\pdfoutput=1
\documentclass[prd,tightenlines,twoside,secnumarabic,floatfix,nofootinbib,10pt]{revtex4}
\usepackage{amsmath,amssymb,bbold,epsfig}
\usepackage{amsfonts}
\usepackage{slashed}
\usepackage{dsfont}
\usepackage{graphicx,color}
\usepackage{bbm}
\usepackage[usenames,dvipsnames]{xcolor}
\definecolor{lightgray}{gray}{0.90}
\usepackage[colorlinks=true,linkcolor=red,citecolor=green,urlcolor=blue,bookmarks=true,bookmarksopen=true,pdfpagemode=None,pdfstartview=FitH]{hyperref}

\newcommand{\be}{\begin{equation}}
\newcommand{\ee}{\end{equation}}
\newcommand{\bea}{\begin{eqnarray}}
\newcommand{\eea}{\end{eqnarray}}
\begin{document}

\title{
Atmospheric neutrinos, $\nu_e$-$\nu_s$ oscillations 
and a novel neutrino evolution equation
}
\author{Evgeny~Akhmedov}
  \altaffiliation{Also at the NRC Kurchatov Institute, Moscow, Russia}
                                   \email[Email: ]{akhmedov@mpi-hd.mpg.de}
\affiliation{
  Max Planck-Institut f\"ur Kernphysik, Saupfercheckweg 1, 69117
       Heidelberg, Germany}

\vspace*{3.5mm}
\date{\today}  
\thispagestyle{empty}
\vspace{-0.8cm}
\begin{abstract}
If a sterile neutrino $\nu_s$ with an eV-scale mass and a sizeable 
mixing to the electron neutrino exists, as indicated by the reactor and 
gallium neutrino anomalies, a strong resonance enhancement of 
$\nu_e$-$\nu_s$ oscillations of atmospheric neutrinos should occur in 
the TeV energy range. At these energies neutrino flavour transitions in 
the 3+1 scheme depend on just one neutrino mass squared difference and 
are fully described within a 3-flavour oscillation framework. We 
demonstrate that the flavour transitions of atmospheric $\nu_e$ can 
actually be very accurately described in a 2-flavour framework, with 
neutrino flavour evolution governed by an inhomogeneous 
Schr\"{o}dinger-like equation. Evolution equations of this type have not 
been previously considered in the theory of neutrino oscillations.
\end{abstract}
\vspace{0.2cm}
\vspace{0.3cm}

\maketitle
\section{\label{sec:intro}Introduction}
A number of anomalies observed in short baseline neutrino experiments can be 
explained through the existence of a sterile neutrino $\nu_s$ with mass in the eV range 
\cite{LSND1,LSND2,MB,React1,React2,React3,gallium1,gallium2,gallium3,gallium4,
gallium5,gallium6} (see Refs. \cite{rev1,rev2,rev3,rev4,rev5} for reviews).  
In particular, the so-called reactor neutrino anomaly \cite{React1,React2,
React3} and gallium anomaly \cite{gallium1,gallium2,gallium3,gallium4,
gallium5,gallium6} 
indicate that such a neutrino may have a sizeable mixing with electron 
neutrinos, $|U_{e4}|\sim 0.1$. Though the evidence for 
the existence of 
an eV-mass sterile neutrino is not compelling,  
this is a very exciting possibility, and 
a considerable amount of work, both theoretical and experimental, has 
been and is being done 
to explore it. 
In particular, a number of experiments focussed 
on searches of an eV-scale sterile neutrino  
are currently being performed or planned. 

One possible way of looking for a light sterile neutrino is to study the 
earth matter effects on 
neutrino oscillations in atmospheric 
and long baseline accelerator 
experiments. 
In particular, 
while $\nu_\mu\leftrightarrow \nu_\tau$ atmospheric neutrino oscillations 
are in general only very weakly 
affected by the earth matter, 
$\nu_\mu\leftrightarrow \nu_s$ oscillations may be strongly influenced by it, 
leading to characteristic energy and zenith angle dependent distortions of the 
observed muon neutrino flux (see 
\cite{ALL,
a2,a3,a4,a5,a6,a7,a8,a9,a10,a11,a12,a13,a14,a15,a16,a17,a18,a19,a20,a21,a22,
a23,a24,a25,a26,a27,IceCube1,a30} for an incomplete list 
of references). Oscillations between $\nu_e$ and $\nu_s$ may also be strongly 
affected by the earth's matter.

The simplest framework to describe active-sterile neutrino oscillations is 
the so-called 3+1 scheme, with three 
light mass-eigenstate neutrinos consisting predominantly of the 
usual active neutrinos $\nu_e$, $\nu_\mu$, $\nu_\tau$ and a heavier 
(or much lighter) state $\nu_4$, consisting mainly of a sterile state 
$\nu_s$. The 
short baseline anomalies mentioned above together with constraints 
from cosmology suggest that $\nu_4$ is the heavier state, with a mass 
$m_4={\cal O}(1)$ eV. 
In this case in the energy range of a few TeV one can expect resonantly 
enhanced $\bar{\nu}_\mu\leftrightarrow \bar{\nu}_s$ and  
$\nu_e\leftrightarrow \nu_s$ oscillations of atmospheric neutrinos 
that come to detectors  
from the lower hemisphere and therefore 
pass through the matter of the 
earth. Disappearance of atmospheric $\bar{\nu}_\mu$ due to oscillations 
into a sterile state driven by $\Delta m^2\sim 1$ eV$^2$ was considered in 
\cite{a17,a19,a21,a22,a23,a24,a25,a26,a27}, and 
the recent analysis of the IceCube data \cite{IceCube1} has put stringent 
constraints on the corresponding allowed parameter space. 
The matter-enhanced 
$\nu_e$-$\nu_s$ oscillations of atmospheric neutrinos have been considered 
in \cite{a25}, where 
it was pointed out that the available IceCube data posed only rather weak 
constraints 
on the parameters governing this oscillation channel.

In the present paper we concentrate on $\nu_e$-$\nu_s$ oscillations of 
atmospheric neutrinos. As 
mentioned above, they are expected to be 
strongly enhanced by the earth's matter in the TeV energy range, making them 
an interesting 
candidate for discovery of sterile neutrinos. 
For baselines limited by the diameter of the earth and TeV-scale neutrtino 
energies, neutrino flavour transitions in the 3+1 scheme 
are governed 
by just one neutrino mass squared difference, $\Delta m_{41}^2$, and are fully 
described within 3-flavour oscillation frameworks \cite{a21,a26}. We 
demonstrate that the flavour transitions of atmospheric $\nu_e$ 
can actually be very accurately described in 
a 2-flavour approach, 
where neutrino oscillations are governed by 
an inhomogeneous Schr\"{o}dinger-like equation. 
Evolution equations of this type have never been previously used for describing 
neutrino oscillations.  

While numerical integration of the full 3-flavour neutrino evolution equation 
does not in general pose any calculational difficulties, the 2-flavour approach 
is more advantageous in several aspects. In particular, it admits simple 
analytical solutions for a number of matter density profiles (such as 
e.g.\ constant matter density, 
several layers of constant densities and an arbitrary density profile in the 
adiabatic approximation) which are much more transparent and more easily 
tractable than the corresponding 3-flavour solutions.

\section{\label{sec:form}Formalism}

Neutrino oscillations in matter are described in the 3+1 framework by the 
evolution equation 
\be
i\frac{d}{dx}\nu = \left[U{\rm diag}\left(0, \frac{\Delta m_{21}^2}{2E}, 
\frac{\Delta m_{31}^2}{2E}, \frac{\Delta m_{41}^2}{2E}\right)U^\dag +
{\rm diag}[V_e(x), 0, 0, V_n(x)]\right]\nu\,,
\label{eq:evol1}
\ee
where $\nu=(\nu_e, \nu_\mu, \nu_\tau, \nu_s)^T$\!, $U$ is the leptonic mixing 
matrix, $\Delta m_{ik}^2=m_i^2-m_k^2$, $G_F$ is the Fermi constant, and 
the neutrino potentials are 
\be
V_e(x)=\sqrt{2}G_F N_e(x)\,,\qquad\quad
V_n(x)=\sqrt{2}G_F N_n(x)/2\,,\qquad\quad
\label{eq:VeVn}
\ee
with $N_e$ and $N_n$ being the electron and neutron number densities of matter, 
respectively. 

For the TeV energy range and terrestrial baselines 
the energy splittings $\Delta m_{21}^2/2E$ and $\Delta m_{31}^2/2E$ can be 
neglected. 
In this case neutrino oscillations are governed by just one mass squared 
difference, 
\be
\Delta \, \equiv\, \frac{\Delta m_{41}^2}{2E}\,,
\label{eq:delta}
\ee
and CP violating effects in neutrino oscillations are unobservable. 
One can therefore choose the mixing matrix $U$ to be real. 
The evolution equation (\ref{eq:evol1}) then takes the form 
\be
i\frac{d}{dx}
\left(
\begin{array}{c} \nu_e \vspace*{0.5mm}\\ \nu_\mu \vspace*{0.5mm}\\ 
\nu_\tau \vspace*{0.5mm}\\ \nu_s
\end{array}
\right) = 
H \left(
\begin{array}{c} \nu_e\vspace*{0.5mm} \\ \nu_\mu \vspace*{0.5mm}\\ 
\nu_\tau \vspace*{0.5mm}\\ \nu_s
\end{array}
\right) = 
\left(
\begin{array}
{llll}
U_{e4}^2\Delta+V_e & ~U_{e4}U_{\mu4}\Delta & ~U_{e4}U_{\tau4}\Delta & 
~U_{e4}U_{s4}\Delta \vspace*{0.5mm}\\ 
U_{\mu 4}U_{e4}\Delta & ~U_{\mu4}^2\Delta &
~U_{\mu4}U_{\tau4}\Delta  & ~U_{\mu4}U_{s4}\Delta \vspace*{0.5mm} \\
U_{\tau4}U_{e4}\Delta &~U_{\tau4}U_{\mu4}\Delta &
~U_{\tau 4}^2\Delta  & ~U_{\tau4}U_{s4}\Delta \vspace*{0.5mm} \\
U_{s4}U_{e4}\Delta &~U_{s4}U_{\mu4}\Delta &
~U_{s4}U_{\tau4}\Delta  & ~U_{s4}^2\Delta+V_n 
 \end{array}
\right) \left(
\begin{array}{c} \nu_e\vspace*{0.5mm} \\ \nu_\mu \vspace*{0.5mm}\\ 
\nu_\tau \vspace*{0.5mm}\\ \nu_s
\end{array}
\right). 
\label{eq:evol1a}
\ee
{}{From experiment we know that 
$|U_{\mu 4}|,|U_{\tau 4}|\lesssim 0.2$. Unitarity then implies $|U_{s4}|=\sqrt{1-U_{e4}^2-U_{\mu 4}^2-U_{\tau 4}^2}
\simeq 1$. 

It will 
be convenient for us to perform a rotation $\nu_\mu$, $\nu_\tau$ $\to$
$\nu_\mu'$, $\nu_\tau'$, that is, to introduce a new basis for the 
neutrino amplitudes according to 
\be
\nu'=V\nu\,,
\qquad
V=\left(
\begin{array}{cccc}
1& ~ 0& 0&0  \\
0 & c_\beta & s_\beta & 0 \\
0 & -s_\beta & c_\beta & 0 \\
0 & 0 & 0 & 1 \end{array}
\right),
\label{eq:newbasis}
\ee
where $c_\beta\equiv\cos\beta$, $s_\beta\equiv\sin\beta$. The neutrino 
evolution equation 
in the new basis reads $i(d/dx)\nu'=VHV^\dag\nu'$. 
Note that 
the rotation with $V$ does not affect the matrix of neutrino 
potentials, and therefore the evolution equation in the primed basis  
has the same form as eq.~(\ref{eq:evol1a}), except that one has to replace 
$\nu_\mu$ and $\nu_\tau$ by 
\be
\nu_\mu'=c_\beta \nu_\mu+s_\beta \nu_\tau\,,\qquad  
\nu_\tau'=-s_\beta \nu_\mu+c_\beta \nu_\tau\,,
\label{eq:primedAmpl}
\ee
and $U_{\mu 4}$ and $U_{\tau 4}$ in the effective Hamiltonian by, respectively, 
\be
U_{\mu 4}'=c_\beta U_{\mu 4}+s_\beta U_{\tau 4}\,,
\qquad
U_{\tau 4}'=-s_\beta U_{\mu 4}+c_\beta U_{\tau 4}\,.
\label{eq:beta1}
\ee
It is easy to see now that one can reduce the 3+1 neutrino evolution equation 
to a 3-flavour form \cite{a21,a26}. Indeed, we can choose the angle $\beta$ by 
requiring, e.g., $U_{\tau4}'=0$, which gives  
\be
s_\beta=\frac{U_{\tau 4}}{\sqrt{U_{\mu 4}^2+U_{\tau 4}^2}}\,,\qquad
c_\beta=\frac{U_{\mu 4}}{\sqrt{U_{\mu 4}^2+U_{\tau 4}^2}}\,, 
\qquad U_{\mu 4}'=
\sqrt{U_{\mu 4}^2+U_{\tau 4}^2}\,.
\label{eq:sc}
\ee
The condition $U_{\tau4}'=0$ results in 
vanishing the third line and the third column of the effective Hamiltonian in 
the rotated basis $VHV^\dag$. This means that the state 
\be
\nu_\tau'=\frac{1}{\sqrt{U_{\mu4}^2+U_{\tau4}^2}}\big(-U_{\tau4}\nu_\mu+
U_{\mu4}\nu_\tau\big) 
\label{eq:nonosc}
\ee
completely decouples from the rest of the neutrino system and does not 
evolve, whereas the orthogonal combination of $\nu_\mu$ and $\nu_\tau$, 
\be
\nu_\mu'=\frac{1}{\sqrt{U_{\mu4}^2+U_{\tau4}^2}}\big(U_{\mu4}\nu_\mu+U_{\tau4}
\nu_\tau\big),
\label{eq:osc}
\ee
is mixed with $\nu_e$ and $\nu_s$. Thus, in the limit $\Delta m_{21}^2, 
\Delta m_{31}^2\to 0$ neutrino flavour transitions take place only 
between $\nu_e$, $\nu_\mu'$ and $\nu_s$.%
\footnote{Obviously, had we chosen $U_{\mu4}'=0$ instead of 
$U_{\tau4}'=0$, the decoupled state would have been $\nu_\mu'$ 
rather than $\nu_\tau'$. However, the physical content of the 
oscillating and non-oscillating neutrino states in both cases is, of course, 
the same, and is given by the right-hand sides of eqs.~(\ref{eq:osc}) and 
(\ref{eq:nonosc}), respectively.}
The corresponding 3-flavour evolution equation is
\be
i\frac{d}{dx}
\left(
\begin{array}{ccc} \nu_e \\ \nu_\mu' \\ \nu_s
\end{array}
\right) = 
H'\!\left(
\begin{array}{ccc} \nu_e \\ \nu_\mu' \\ \nu_s
\end{array}
\right),  
\label{eq:evol2}
\ee
where 
\be
H' \,\equiv\, 
\left(
\begin{array}{ccc}
H_{ee} & H_{e\mu}' & H_{es}\\H_{e\mu}' &H_{\mu\mu}' & H_{\mu s}' \\
H_{es}& H_{\mu s}' & H_{ss} 
\end{array}
\right) \,=\,
\left(
\begin{array}{lll}
U_{e4}^2\Delta+V_e(x)& U_{e4}U_{\mu 4}'\Delta & 
\;\,U_{e4}U_{s4} \Delta \vspace*{0.5mm}\\
U_{e4}U_{\mu4}'\Delta & U_{\mu 4}'^2\Delta & 
\;\,U_{\mu 4}'U_{s4} \Delta \vspace*{0.5mm}\\
U_{e4}U_{s4}\Delta & U_{\mu 4}'U_{s4}\Delta & 
\;\,U_{s4}^2\Delta + V_n(x) 
\end{array}
\right).
\label{eq:H1}
\ee

We are interested in the $\nu_e \leftrightarrow 
\nu_s$ oscillations inside the earth at the energies close to the MSW 
resonance energy in this channel. The transitions in the $\nu_\mu'-\nu_e$ 
and $\nu_\mu'-\nu_s$ channels are then non-resonant, so that the 
probabilities of $\nu_\mu'$ oscillations remain small. However, at the 
energies of interest the original (unoscillated) flux of $\nu_\mu$ is 
significantly larger than the original $\nu_e$ flux ($F_\mu^{(0)}\sim 20 
F_e^{(0)}$), and the same holds true for the $\nu_\mu'$ flux,%
\footnote{By which we mean the weighted sum of the original $\nu_\mu$ and
$\nu_\tau$ fluxes, 
$F_{\mu'}^{(0)}=\big(|U_{\mu4}|^2F_\mu^{(0)}+|U_{\tau4}|^2F_\tau^{(0)}\big)/
\big(|U_{\mu4}|^2+|U_{\tau4}|^2\big)$. 
Note that at the TeV energy scale the flux 
of atmospheric $\nu_\tau$ is still rather small, $F_\tau^{(0)}\sim 1\%
F_\mu^{(0)}$. 
}  
barring the possibility $|U_{\mu4}|\ll|U_{\tau4}|$. 
Therefore, in calculating the $\nu_e$ flux at the detector site, one cannot 
in general neglect the transitions from $\nu_\mu'$, unless the mixing 
parameter $U_{\mu4}'$ satisfies $|U_{\mu4}'|\ll |U_{e4}|$. Still, a 
significant simplification is possible if one takes into account that, 
though the transitions $\nu_\mu'\to \nu_e$ and $\nu_\mu'\to \nu_s$ have to be 
taken into account  
because of the large initial $\nu_\mu'$ flux,  
the inverse processes, i.e.\ back reaction of $\nu_e$ and $\nu_s$ on the 
$\nu_\mu'$ flux, can be neglected. 
This means that the flux of $\nu_\mu'$ can be considered as fixed, serving 
as an external source for the evolution equations for the $\nu_e$ and 
$\nu_s$ amplitudes. The 3-flavour oscillation problem of 
eqs.~(\ref{eq:evol2}), (\ref{eq:H1}) can therefore be reduced to a much 
simpler 2-flavour one.

The argument proceeds as follows. First, since the initial value 
$|\nu_\mu'(0)|$ is much larger than $|\nu_e(0)|$ and $\nu_s(0)=0$, 
we retain on the right-hand side of the equation for the amplitude 
$\nu_\mu'(x)$ in eq.~(\ref{eq:evol2}) only the term $H_{\mu\mu}'\nu_\mu'(x)$. 
The equation can then be immediately solved, giving 
\be
\nu_\mu'(x)=\nu_\mu'(0)e^{-i H_{\mu\mu}' x}\,.
\label{eq:IC1}
\ee
Substituting this into the equations for the amplitudes $\nu_e(x)$ and 
$\nu_s(x)$ yields an inhomogeneous 2-flavour evolution equation:
\be
i\frac{d}{dx}
\left(\!
\begin{array}{c} \nu_e \\ \nu_s
\end{array}
\!\right) = 
\left(\!
\begin{array}{cc}
H_{ee} & H_{es} \\
H_{es} & H_{ss} \end{array}
\right) \left(
\begin{array}{c} \nu_e \\ \nu_s
\end{array}
\!\right) + \left(\!\begin{array}{c} f_e \\ f_s\end{array}
\!\right). 
\label{eq:evol3}
\ee
Here the external sources are 
\be
f_e(x)=H_{e\mu}' \nu_\mu'(0)e^{-i H_{\mu\mu}' x}\,, \qquad 
f_s(x)=H_{\mu s}' \nu_\mu'(0)e^{-i H_{\mu\mu}' x}\,. 
\label{eq:sources}
\ee
Eq.~(\ref{eq:evol2}) should be solved with the initial conditions 
\be
\nu_e(0)=1\,,\qquad \nu_s(0)=0\,.
\label{eq:IC2}
\ee
This would be equivalent to solving the original system (\ref{eq:evol2}) 
with the initial \vspace*{1.5mm} conditions%
\footnote{Here we have taken into account eq.~(\ref{eq:osc}) and have 
neglected the small initial flux of atmospheric $\nu_\tau$.
}
\be
\nu_e(0)=1\,,\qquad\quad
\nu_\mu'(0)=\frac{U_{\mu4}}{\sqrt{U_{\mu4}^2+U_{\tau4}^2}}
\big(F_\mu^{(0)}/F_e^{(0)}\big)^{1/2}\,,\qquad\quad \nu_s(0)=0\,
\label{eq:IC3}
\ee
\noindent	
and with the back reaction on $\nu_\mu'(x)$ neglected. Such a calculation 
would have been correct if the initial neutrino state were a coherent 
superposition of $\nu_e$ and $\nu_\mu$, whereas in reality the 
original unoscillated atmospheric neutrino flux consists of an incoherent 
sum of the $\nu_e$ and $\nu_\mu$ fluxes. This can be taken into account by 
introducing a random phase for $\nu_\mu'(0)$ in eq.~(\ref{eq:IC3}) 
according to $\nu_\mu'(0)\to e^{i\varphi}\nu_\mu'(0)$ 
and averaging over the random phase $\varphi$ 
at the probabilities level. Such an averaging can be readily 
done, as will be discussed below.

Let us now demonstrate how one can find the solution $(\nu_e, \nu_s)^T$ of 
eq.~(\ref{eq:evol3}) provided that the solution $(\nu^{(0)}_e, \nu^{(0)}_s)^T$ 
of the corresponding homogeneous equation (which has the form of the standard 
2-flavour evolution equation for neutrino oscillations in matter) is known. 
To this end, it will prove to be convenient to subtract from the effective 
Hamiltonian in (\ref{eq:evol3}) the term (1/2)$(H_{ee}+H_{ss})\!\cdot
\!\!\mathbbm{1}$, rendering the resulting Hamiltonian traceless. 
Simultaneously, the external sources $f_e(x)$ and $f_s(x)$ should be rephased 
according to
\be
f_{e,s}(x)\to f_{e,s}'(x)=f_{e,s}(x)e^{\frac{i}{2}\int_0^x 
A(x')dx'}\,,\quad
A(x)\equiv H_{ee}(x)+H_{ss}(x)\,.
\label{eq:rephase}
\ee
The amplitudes $\nu_e(x)$ and $\nu_s(x)$ should be rephased similarly, 
but this would not affect the oscillation probabilities and therefore 
we keep for these amplitudes the old (unprimed) notation. Thus, we 
have to solve the evolution equation 
\be
i\frac{d}{dx}
\left(\!
\begin{array}{c} \nu_e \\ \nu_s
\end{array}
\!\right) = 
\left(\!
\begin{array}{cc}
\frac{H_{ee}-H_{ss}}{2} & H_{es} \\
H_{es} & \frac{H_{ss}-H_{ee}}{2} \end{array}
\right) \left(\begin{array}{c} \nu_e \\ \nu_s
\end{array}
\!\right) + \left(\!\begin{array}{c} f_e' \\ f_s'\end{array}
\!\right)  
\label{eq:evol4}
\ee
with the initial condition (\ref{eq:IC2}). 

Assume that we know the solution $(\nu_e^{(0)}, \nu_s^{(0)})^T$ of 
the homogeneous equation, (i.e.\ of eq.~(\ref{eq:evol4}) with $f_e'=f_s'=0$) 
with the initial condition $(1, 0)^T$. 
We denote it 
\be
\nu^{(0)}(x)=
\left(\!
\begin{array}{c} 
\nu_e^{(0)}(x)\\ \nu_s^{(0)}(x)
\end{array}
\!\right) \equiv
\left(\!
\begin{array}{r} 
\alpha(x,0) \\ 
-\beta^*(x,0) 
\end{array}
\right).
\label{eq:sol1}
\ee
Then the quantity $-i\sigma_2 \nu^{(0)*}(x)=(\beta(x,0), \alpha^*(x,0))^T$ 
is also a solution of the same equation, but satisfying the initial condition 
$(0, 1)^T$.%
\footnote{Note that this is only true in the case of traceless 2-flavour 
Hamiltonians.} 
It is then easy to see that the matrix $S(x,0)$ with the columns given by 
the two independent solutions of the homogeneous equation, $\nu^{(0)}(x)$ 
and $-i\sigma_2 \nu^{(0)*}(x)$, \vspace*{-1.5mm} i.e.\
\be
S(x,0)=\left(
\begin{array}{cc} \alpha(x,0) & \beta(x,0) \\ -\beta^*(x,0) &\alpha^*(x,0)
\end{array}\right),
\label{eq:S}
\ee
is actually the evolution matrix of the  homogeneous equation. Indeed, it 
satisfies the 
equation $i(d/dx)S(x,0)=H^{\rm (2f)}S(x,0)$ (where $H^{\rm (2f)}$ is the 
Hamiltonian of eq.~(\ref{eq:evol4})) and obeys the initial condition 
$S(0,0)=\mathbbm{1}$. 
Note that the matrix $S$ is unitary, as it should be.  

The solution of the full inhomogeneous equation~(\ref{eq:evol4}) is then 
readily found: One can check by direct substitution \vspace*{-1.5mm} that 
\be
\left(
\begin{array}{c} 
\nu_e(x)\\ \nu_s(x)
\end{array}
\right) = 
\left(
\begin{array}{c} 
\nu_e^{(0)}(x)\\ \nu_s^{(0)}(x)
\end{array}
\right)-i S(x, 0)\int_0^x S^{-1}(x',0)
\left(
\begin{array}{c} 
f_e'(x') \\ 
f_s'(x') 
\end{array}
\right) dx'
\label{eq:sol2}
\vspace*{1mm}
\ee

\noindent
satisfies eq.~(\ref{eq:evol4}) with the initial condition (\ref{eq:IC2}). 
This leads to very simple expressions for $\nu_e(x)$ and $\nu_s(x)$.
In terms of the parameters $\alpha$ and $\beta$ we have 
\begin{align}
\nu_e(x)=\alpha(x,0)-i\int_0^x\left[\alpha(x,0)\alpha^*(x',0)+\beta(x,0)
\beta^*(x',0)\right]f_e'(x')dx' ~~~~~\,
\nonumber \\
+\,i\int_0^x\left[\alpha(x,0)\beta(x',0)-\beta(x,0)\alpha(x',0)\right]
f_s'(x')dx'\,,~~~~~~\,
\label{eq:sol4a}
\\
\nu_s(x)=-\beta^*(x,0)+i\int_0^x\left[\beta^*(x,0)\alpha^*(x',0)-\alpha^*(x,0)
\beta^*(x',0)\right]f_e'(x')dx' 
\nonumber \\
-\,i\int_0^x\left[\alpha^*(x,0)\alpha(x',0)+\beta^*(x,0)\beta(x',0)\right]
f_s'(x')dx'\,.~
\label{eq:sol4b}
\end{align}
These formulas can be further simplified by noting that the expressions in 
the square brackets in (\ref{eq:sol4a}) and (\ref{eq:sol4b}) are actually 
the elements of the matrix $S(x,x')$ (this can also be seen directly from 
eq.~(\ref{eq:sol2})). Indeed, from the well known properties 
of the evolution matrix it follows that $S(x,x')=S(x,0)S(0,x')=S(x,0)
S^{-1}(x',0)$. By making use of eq.~({\ref{eq:S}) one then finds 
\begin{align}
\nu_e(x)&=\alpha(x,0)-i\int_0^x\left[\alpha(x,x')f_e'(x')
+\beta(x,x')f_s'(x')\right]dx'\,,~
\label{eq:sol5a}
\\
\nu_s(x)&={}\!\!-\beta^*(x,0)+i\int_0^x\left[\beta^*(x,x')f_e'(x') 
-\alpha^*(x,x')f_s'(x')\right]dx'\,.~
\label{eq:sol5b}
\end{align}
This is our final result
(though eqs.~(\ref{eq:sol4a}) and (\ref{eq:sol4b}) may also be useful). 
Note that the sources $f_e'(x)$ and $f_s'(x)$ have the same coordinate 
dependence,%
\footnote{Indeed, from eqs.~(\ref{eq:sources}), (\ref{eq:rephase}) 
and (\ref{eq:H1}) one finds $f_s'(x)=(U_{s4}/U_{e4})f_e'(x)$.}
which simplifies the calculation of $\nu_e(x)$ and $\nu_s(x)$. 
Very 
simple analytic expressions for the parameters $\alpha$ and $\beta$
can be obtained, e.g., for constant-density matter or constant-density 
layers model of the earth density profile \cite{param}. 

It is also useful to present expressions (\ref{eq:sol4a}) and 
(\ref{eq:sol4b}) for $\nu_e(x)$ and $\nu_s(x)$ in a more compact form:
\be
\nu_e(x)=\nu_e^{(0)}(x)-i \nu_e^{(0)}(x) a(x) + i\nu_s^{(0)*}(x) b(x)\,,
\label{eq:sol3a}
\ee
\be
\nu_s(x)=\nu_s^{(0)}(x)-i \nu_s^{(0)}(x) a(x) - i\nu_e^{(0)*}(x) b(x)\,,
\label{eq:sol3b}
\ee
where
\begin{align}
a(x)=\int_0^x 
\left[\nu_e^{(0)*}(x')f_e'(x')+\nu_s^{(0)*}(x')f_s'(x')\right] dx'\,,
\nonumber \\ 
b(x)=\int_0^x 
\left[-\nu_s^{(0)}(x')f_e'(x')+\nu_e^{(0)}(x')f_s'(x')\right] dx'\,.
\label{eq:ab}
\end{align}

In order to obtain the survival probability of electron neutrinos 
$\bar{P}_{ee}^{(2f)}(x)$ we have to average $|\nu_e(x)|^2$, where $\nu_e(x)$ is 
the solution of the 2-flavour evolution equation (\ref{eq:evol4}) (or 
equivalently of eq.~(\ref{eq:evol3})), over the 
random phase $\varphi$ of the initial amplitude $\nu_\mu'(0)$. 
The quantities $a(x)$ and $b(x)$ are proportional to $\nu_\mu'(0)$ (because so 
are $f_e'(x)$ and $f_s'(x)$), and therefore the averaging over the random 
phase of $\nu_\mu'(0)$   
is achieved by 
discarding the contribution to $|\nu_e(x)|^2$ coming from the interference 
of the first term on the right-hand side of eq.~(\ref{eq:sol3a}) with the 
remaining two terms:%
\footnote{Indeed, since the last two terms on the right hand side of 
eq.~(\ref{eq:sol3a}) are proportional to $e^{i\varphi}$, they can be written 
as $\nu_e(x)-\nu_e^{(0)}(x)=e^{i\varphi}B(x)$, where $B(x)$ is independent of 
$\varphi$. Then the squared modulus of eq.~(\ref{eq:sol3a}) yields 
$|\nu_e^{(0)}(x)|^2+|B(x)|^2+2\cos\varphi{\rm Re}[\nu_e^{(0)}(x)^*B(x)]
-2\sin\varphi{\rm Im}[\nu_e^{(0)}(x)^*B(x)]$. Upon averaging over the random 
phase $\varphi$ the last two terms in this expression vanish. Taking into 
account that $|B(x)|=|\nu_e(x)-\nu_e^{(0)}(x)|$ yields eq.~(\ref{eq:Peff}).
}
\be
\bar{P}_{ee}^{(2f)}(x)=|\nu_e^{(0)}(x)|^2+|\nu_e(x)-\nu_e^{(0)}(x)|^2\,.
\label{eq:Peff}
\ee
If $\nu_e(x)$ is found from eq.~(\ref{eq:sol5a}), the averaging would 
correspond to neglecting the interference between the first and the second 
terms on the right-hand side of this equation when calculating $|\nu_e(x)|^2$. 
The oscillated $\nu_e$ flux at the detector site (corresponding to the baseline 
$L$) is then 
\be
F_e=\bar{P}_{ee}^{(2f)}(L) F_e^{(0)}\,. 
\label{eq:N_e4}
\ee
The dependence on the original muon neutrino flux $F_\mu^{(0)}$ is implicitly 
contained in $\bar{P}_{ee}^{(2f)}(L)$ through the initial 
conditions (\ref{eq:IC3}).  

At the same time, when the oscillated neutrino flux is calculated directly 
from the 3-flavour neutrino evolution, without approximating it by a 
2-flavour one, the electron neutrino flux at the detector site is given by  
\be
F_e=P_{ee}(L) F_e^{(0)}+P_{\mu' e}(L) F_{\mu'}^{(0)}\,,
\label{eq:N_e1}
\ee
where the oscillation probabilities $P_{ee}(L)$ and $P_{\mu' e}(L)$ should 
be obtained 
from the evolution equation~(\ref{eq:evol2}).

\section{\label{sec:results}Results}

In this section we compare the 
results of the complete 3-flavour calculations with those obtained in 
the approximate 2-flavour framework. 
In both cases constant density layers model was 
used for the matter density profile of the earth (see \cite{param} for 
the numerical values of the parameters used and for the explicit formula for 
the 2-flavour evolution matrix $S$). 
For completeness, we quote here the 2-flavour expressions for the 
parameters $\alpha(x,0)$ and $\beta(x,0)$ for the three-layer model of the 
earth's density profile for neutrinos crossing the core of the earth 
(mantle-core-mantle trajectories) \cite{param}:
\be
\alpha(x,0)=Y-iX_3\,,\qquad\quad \beta(x,0)=-i(X_1-iX_2)\,,
\ee
where 
\bea
&Y=c_1 c_2 - s_1 s_2 \cos(2\theta_1-2\theta_2)\,,\qquad
&X_1=s_1 c_2\sin 2\theta_1+s_2 c_1\sin 2\theta_2\,,
\nonumber \\
&X_2=-s_1 s_2 \sin(2\theta_1-2\theta_2)\,,\qquad\qquad
&X_3=-(s_1 c_2\cos 2\theta_1+s_2 c_1\cos 2\theta_2)\,. 
\eea
Here $\theta_1$ and $\theta_2$ are the in-matter mixing angles in the mantle 
and in the core of the earth, respectively; $s_1$ ($c_1$) is the sine (cosine) 
of the oscillation phase in each mantle layer, while $s_2$ ($c_2$) 
is the sine (cosine) of the oscillation phase in earth's core. For mantle-only 
crossing neutrino trajectories one has to put $\theta_2=\theta_1$ and replace 
$c_1 c_2-s_1 s_2$ and $s_1 c_2+s_2 c_2$ by, respectively, the cosine and sine 
of the total oscillation phase.

In figs.~\ref{fig:1} and \ref{fig:2} the solid (red) curves 
show the results of the full 3-flavour calculations, whereas the 
dashed (blue) curves represent the results obtained in the 2-flavour 
approach with averaging over the random phase of the initial $\nu_\mu'$ 
state, as discussed above. We observe a very good agreement between the 
results of these two approaches -- the oscillated $\nu_e$ fluxes obtained 
in the 3-flavour and 2-flavour frameworks are almost indistinguishable. 
With increasing $|U_{\mu 4}'|$ the difference between the 2-flavour and 
3-flavour results 
becomes more pronounced, 
compare figs.~\ref{fig:1} and~\ref{fig:2}. 
This is because the approximation of neglecting the back reaction of 
$\nu_e$ and $\nu_s$ on the $\nu_\mu'$ flux becomes less accurate 
in this case. Also the dip at $E\simeq 8$ TeV becomes less deep due to the 
increased $\nu_\mu'\to \nu_e$ transitions.  

For illustration, we also show the $\nu_e$ flux obtained 
from the 2-flavour calculation 
without averaging over the random phase of $\nu_\mu'(0)$ (dot-dashed green 
curves). As expected, this gives a wrong result.
\begin{figure}[h]
  \begin{center}
\includegraphics[width=8.15cm,height=6.5cm]{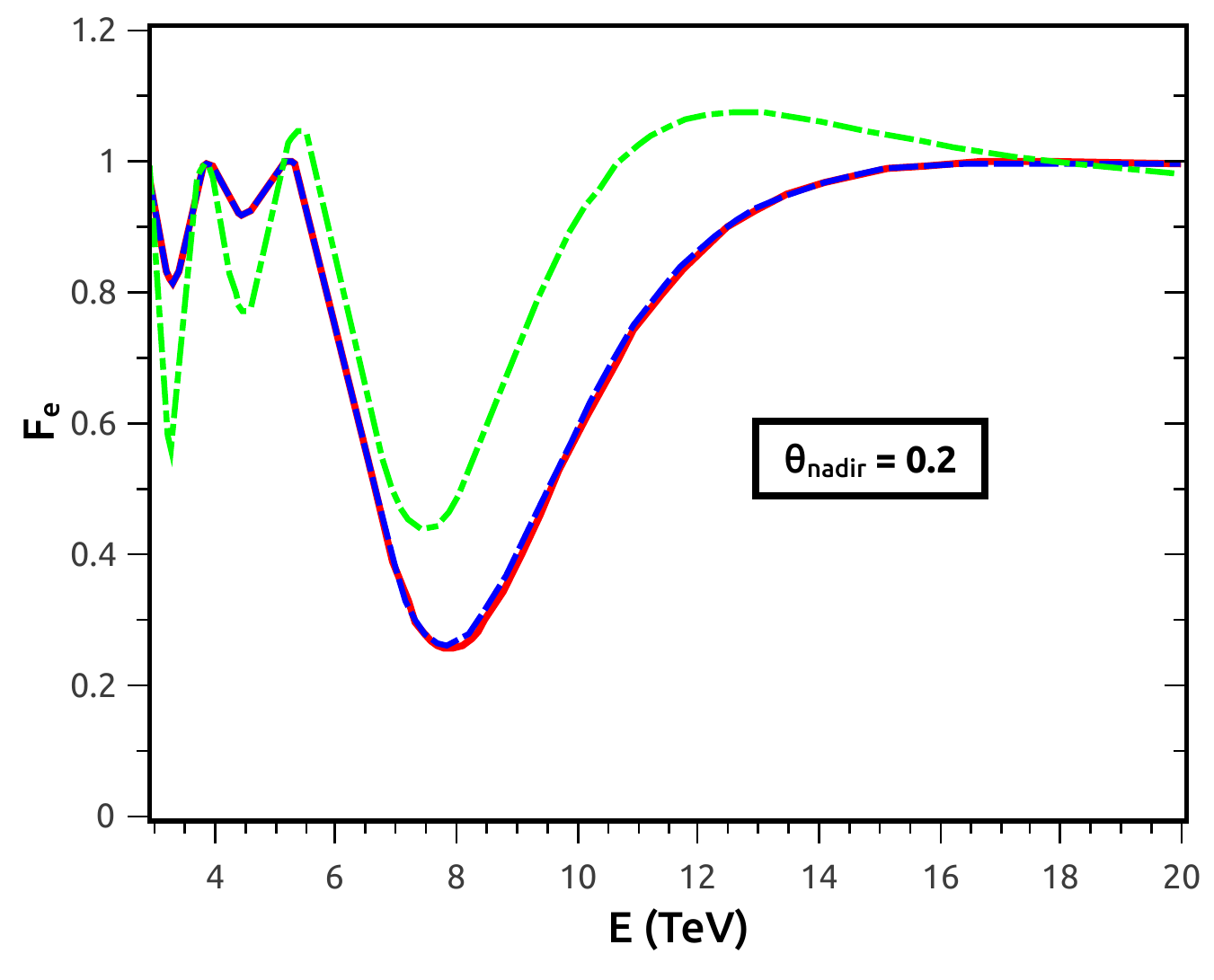}
\includegraphics[width=8.15cm,height=6.5cm]{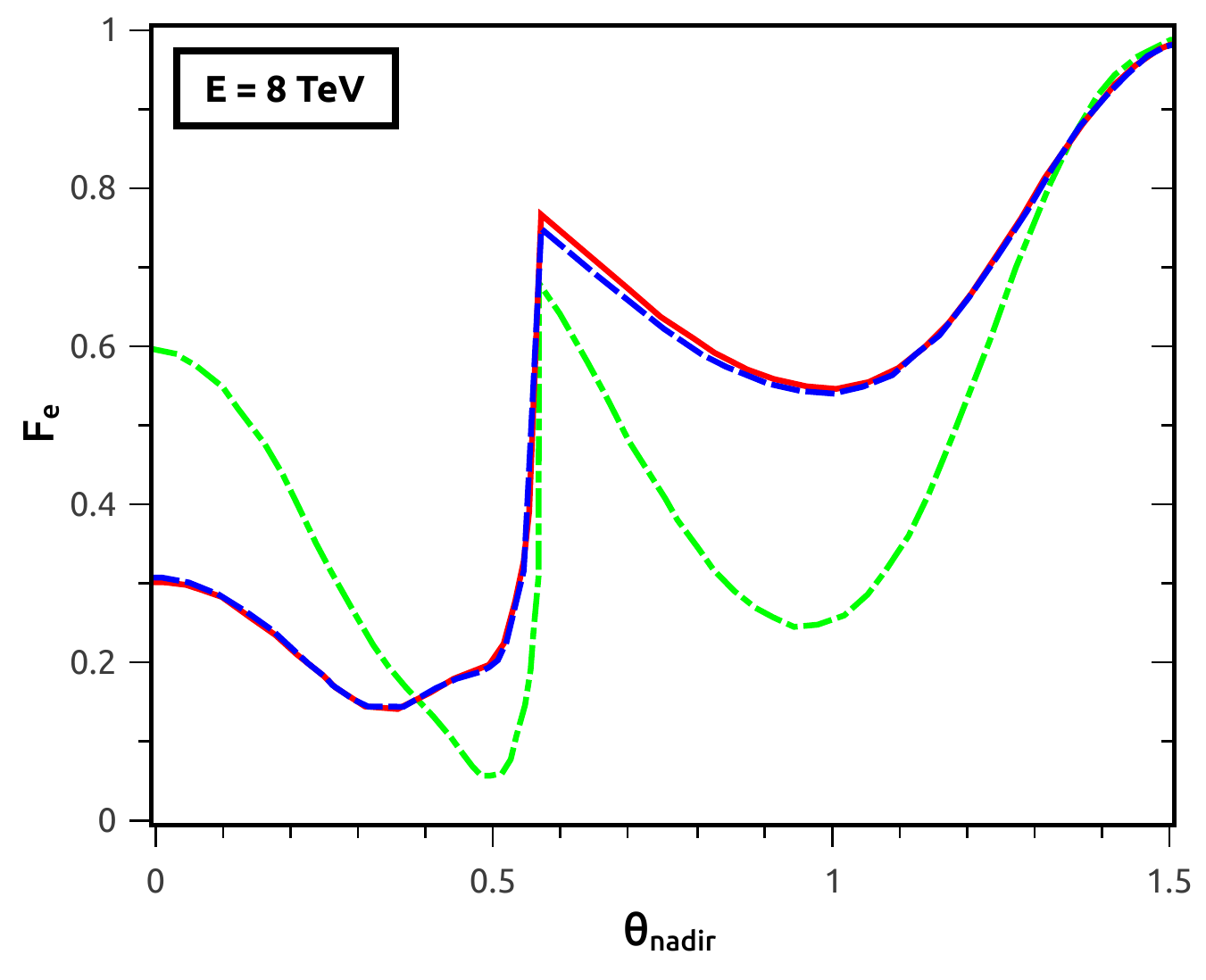}
  \end{center}
\caption{Electron neutrino flux $F_e$ at the detector site (normalized 
to the unit unoscillated flux, $F_e^{(0)}=1$) as a function of the neutrino 
energy $E$ (left panel, nadir angle $\theta_{nadir}=0.2$) and of the nadir 
angle $\theta_{nadir}$ 
(right panel, $E=8$ TeV). Solid (red) curves: full 3-flavour calculation 
[eqs.~(\ref{eq:evol2}), (\ref{eq:N_e1})]. Dashed (blue) curves: 
result based on 
the effective 2-flavour evolution 
[eqs.~(\ref{eq:evol4}), (\ref{eq:N_e4})]. For comparison, dash-dotted (green) 
curves show  the (incorrect) 2-flavour results   
obtained without averaging over the random phase 
of $\nu_\mu'(0)$ (see the text). The following values of parameters 
used: $\Delta m_{41}^2=2.5$ eV$^2$; $U_{e4}=0.2$; 
$U_{\mu 4}'\equiv\sqrt{U_{\mu 4}^2+U_{\tau 4}^2}=0.1$; 
$|\nu_\mu'(0)|=\sqrt{20}\,\nu_e(0)$. 
Constant density layers model of the earth density profile. 
}
\label{fig:1}
\end{figure}
\begin{figure}[!h]
  \begin{center}
\includegraphics[width=8.15cm,height=6.5cm]{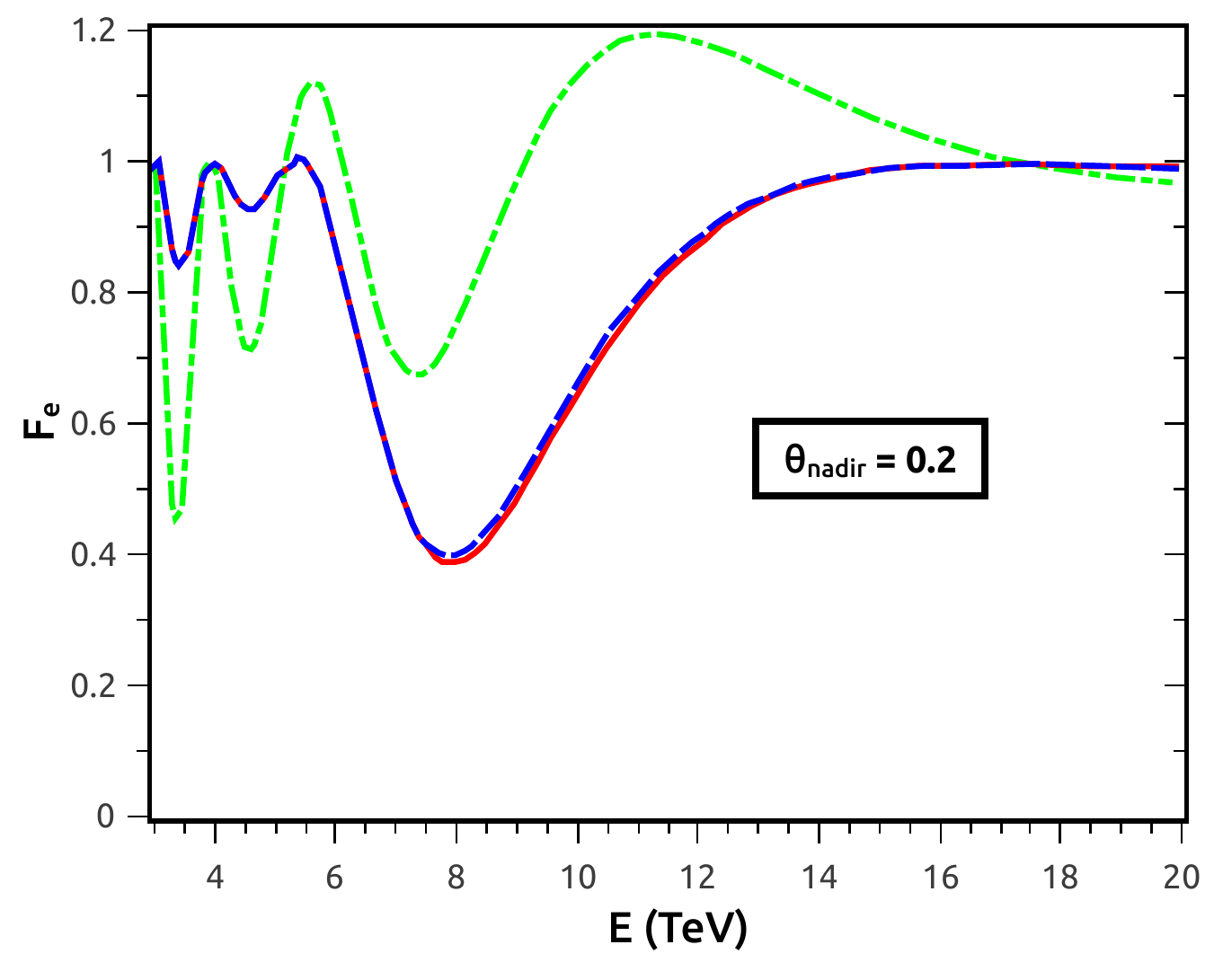}
\includegraphics[width=8.15cm,height=6.5cm]{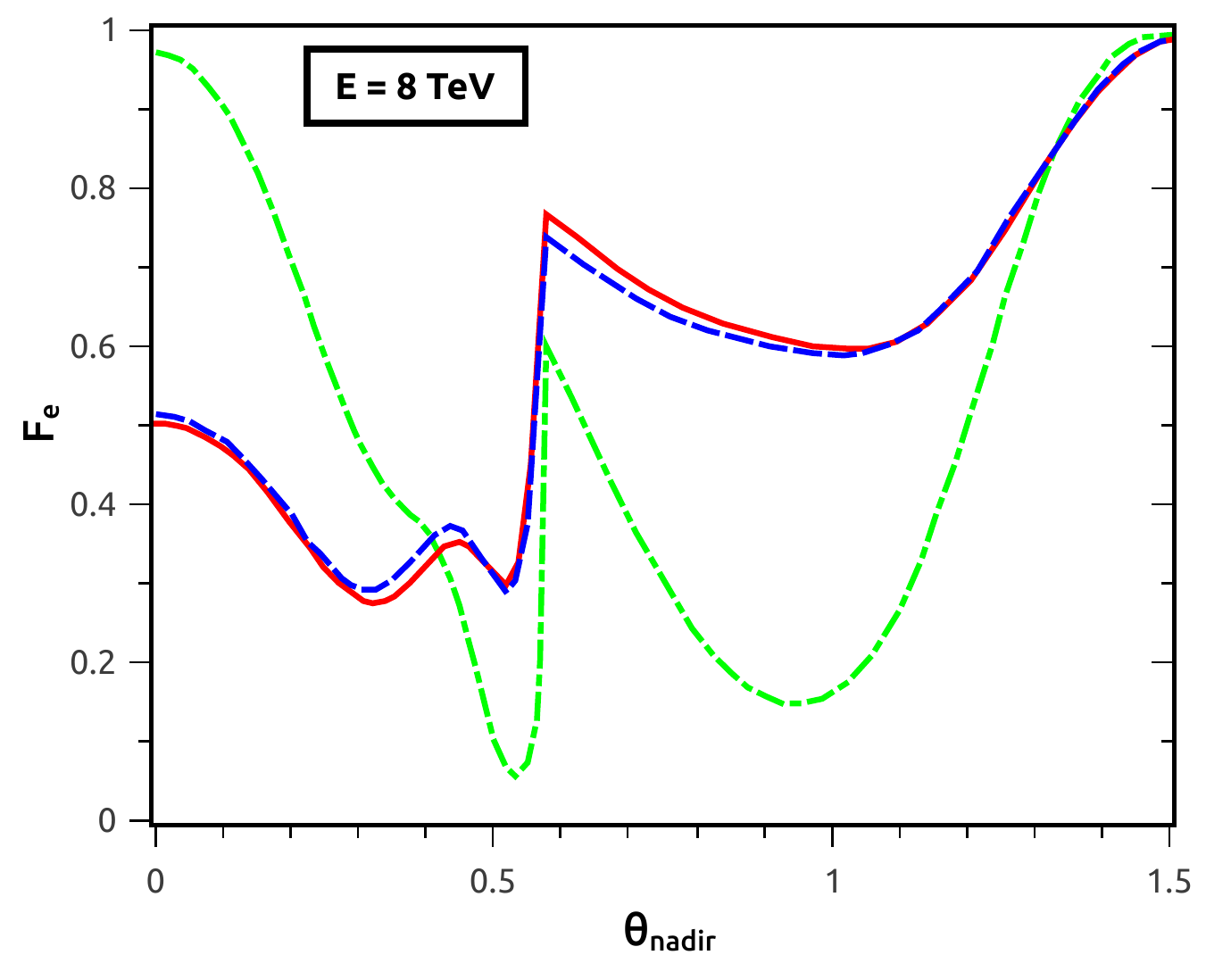}
  \end{center}
\caption{Same as in fig.~\ref{fig:1}, but for 
$U_{\mu4}'=0.15$.}
  \label{fig:2}
\end{figure}

Atmospheric neutrino oscillations in the 3+1 scheme can in principle both 
enhance and suppress the $\nu_e$ flux compared to the unoscillated flux 
$F_e^{(0)}$. Interestingly, our calculations presented in figs.~\ref{fig:1} 
and~\ref{fig:2} 
demonstrate only the reduction of the $\nu_e$ flux. This 
happens because in the considered TeV energy region the $\nu_e\to \nu_s$ 
disappearance is strongly enhanced by matter effects and dominates over 
the $\nu_e$ appearance due to the $\nu_\mu'\to \nu_e$ transitions. 

\section{\label{sec:disc}Summary and discussion}
If a sterile neutrino with an eV-scale mass exists and has a sizeable 
mixing to $\nu_e$, as the reactor and gallium neutrino anomalies suggest, 
the flux of atmospheric $\nu_e$ in the TeV energy region can be 
significantly affected by $\nu_e\leftrightarrow \nu_s$ oscillations. 
Strong enhancement of the 
oscillation probability in this channel can occur due to the MSW and 
parametric resonances of neutrino oscillations in the earth, just like it 
occurs in the GeV energy range for the usual $\nu_e\leftrightarrow \nu_\mu$ and 
$\nu_e\leftrightarrow \nu_\tau$ oscillations driven by the atmospheric mass 
squared difference $\Delta m_{31}^2$ and the mixing angle $\theta_{13}$ (see, 
e.g., \cite{AMS1}).  

The simplest framework to consider active-sterile neutrino oscillations is 
the 3+1 scheme. For TeV-scale neutrinos and terrestrial baselines, oscillations 
in this scheme are governed by just one mass squared difference, 
$\Delta m_{41}^2$, whereas the effects of $\Delta m_{31}^2$ and  
$\Delta m_{21}^2$ can be neglected. In this approximation neutrino oscillations 
in the 3+1 scheme can be reduced to pure 3-flavour oscillations between 
$\nu_e$, $\nu_s$, and $\nu_\mu'$, which is a linear combination of $\nu_\mu$ 
and $\nu_\tau$.  We have demonstrated that in the TeV energy range flavour 
transitions of atmospheric neutrinos can actually 
be very well described by a 2-flavour evolution equation. 
This proved to be possible because the probabilities of oscillations of the 
muon neutrinos 
are rather small in this energy region. This does not mean that the 
oscillations of the atmospheric $\nu_\mu$ can be completely ignored, as their 
inital flux is about a factor of 20 larger than the $\nu_e$ flux in the TeV 
energy region. Therefore, the transitions $\nu_\mu\to\nu_e$ and 
$\nu_\mu\to\nu_s$ should be taken into account. However, the back reaction of 
$\nu_e$ and $\nu_s$ on the $\nu_\mu$ flux can to a good accuracy be neglected. 
As a result, the flux of the muon neutrinos remains practically unchanged and 
serves as an external source for a 2-flavour evolution equation.  
 
It can be seen from figs.~\ref{fig:1} and \ref{fig:2} that our effective 
2-flavour approach provides a very good approximation for the 
full 3-flavour evolution of atmospheric neutrinos in the TeV energy 
range. The main advantage of the 2-flavour framework is that for many matter 
density profiles of practical interest it allows analytical solutions which are 
much simpler and much more transparent than the corresponding 3-flavour 
expressions.  The 2-flavour neutrino evolution equation derived 
here, eq.~(\ref{eq:evol3}) [or eq.~(\ref{eq:evol4})], has a form of an 
inhomogeneous Schr\"{o}dinger-like equation (equation with external sources). 
To the best of the present author's knowledge, evolution 
equations of this type have never been previously used for describing neutrino 
oscillations.

\bigskip
{\em Acknowledgments.} The author is grateful to Alexei Smirnov for useful 
discussions, to Lisa Michaels for checking the numerical results presented in 
Section~\ref{sec:results} and to Farinaldo Queiroz for his help with the 
figures.

\end{document}